\documentclass[12pt]{article}
\usepackage{graphicx}

\input epsf
\def\beq{\begin{eqnarray}}
\def\eeq{\end{eqnarray}}

\def\k{{\bf k}}

\def\lsim{\mathrel{\rlap{\lower3pt\hbox{\hskip0pt$\sim$}}
     \raise1pt\hbox{$<$}}}         %less than or approx. symbol
\def\gsim{\mathrel{\rlap{\lower4pt\hbox{\hskip1pt$\sim$}}
     \raise1pt\hbox{$>$}}}         %greater than or approx. symbol

\usepackage{amsmath}
\usepackage{amsfonts}
\usepackage{verbatim}

\begin{document}

\begin{titlepage}

\thispagestyle{empty}

\begin{flushright}
{NYU-TH-10/02/62}
\end{flushright}
\vskip 0.9cm

\centerline{\Large \bf Field Theory for a Deuteron Quantum Liquid}
\vskip 0.2cm
\centerline{\Large \bf }                    

\vskip 0.7cm
\centerline{ \large Lasha Berezhiani, Gregory Gabadadze and David Pirtskhalava}
\vskip 0.3cm
\centerline{\em Center for Cosmology and Particle Physics,  
Department of Physics,}
\centerline{\em New York University, New York, 
NY  10003, USA}

\vskip 1.9cm

\begin{abstract}

Based on general symmetry principles we study 
an effective Lagrangian for a neutral system 
of condensed spin-1 deuteron nuclei and electrons, 
at greater-than-atomic but less-than-nuclear densities.
We expect such matter to be present in thin layers within 
certain low-mass brown dwarfs.  It may also be produced in future
shock-wave-compression experiments as an effective fuel 
for laser-induced nuclear fusion. We find a  background solution 
of the effective  
theory describing a net spin zero condensate of deuterons 
with their spins aligned and anti-aligned in a certain spontaneously 
emerged preferred direction. The spectrum of low energy collective excitations 
contains  two  spin-waves with linear dispersions -- 
like in antiferromagnets -- as well as gapped longitudinal and 
transverse modes related to  the Meissner effect -- like in superconductors.
We show that counting of the Nambu-Goldstone modes of spontaneously broken 
internal and space-time symmetries obeys,  in a nontrivial way,  
the rules of the Goldstone  theorem for Lorentz non-invariant systems. 
We discuss thermodynamic properties of the condensate,  
and its  potential manifestation in the low-mass brown dwarfs.

\end{abstract}

\vspace{3cm}

\end{titlepage}

\newpage

\begin{center}
{\large \bf 1. Introduction and Summary}
\end{center}

\vskip 0.3cm

The field theory method is an extremely powerful tool for studying  
properties of matter and radiation at various scales. 
The subject of the present work is deuterium matter at 
greater-than-atomic,  but much less-than-nuclear densities. At such  densities
the neutral deuterium atoms are disintegrated into charged 
spin-1 deuteron  nuclei and electrons, with an average inter-particle 
separation being smaller than  $1~ \AA$ but much greater than
$1~fm$.  At temperatures of interest in this work
($T\lsim  10^5~K$), the electrons  form a degenerate Fermi gas.  
Deuterons, on the other hand,  can be regarded  as point-like spin-1 
charged particles, which as argued below, will condense 
beneath certain temperature forming  a quantum liquid.  
A field theory description 
of this condensed state of deuterons is the main 
topic of the present work.

Recently, the condensation of charged 
spin-0 helium-4 nuclei at densities $\rho\sim 10^6~g/cm^3$ 
was studied using the effective field theory method \cite{gr1}. 
Such a dense system  is expected to be 
present in  helium-core white dwarf stars.  It was shown \cite {gp}  that 
these stars  will cool faster because of  the condensation of 
helium-4 nuclei (for a review see, \cite {grrev}).

Here we develop an  effective field theory approach to describe 
the deuteron quantum liquid. This is interesting for a few reasons: 
(i) Densities 
at which deuterons could  condense are much lower than those 
for helium-4 nuclei; as argued below,   at $\rho > 100~g/cm^3$,  the deuteron 
condensate is expected to form upon gradual cooling of the system. 
(ii) The high-density deuteron matter is a promising 
fuel for  laser-induced nuclear fusion.  Hence, there has  been 
a significant experimental  interest in dense deuterium 
(as well as hydrogen) liquids  
(see \cite{ex1} and references therein).
Even though the present-day laboratory densities of shock-wave-compressed 
deuterium matter  are two orders of magnitude below  $100~g/cm^3$, 
the experiments  do see a transition to a liquid metallic state in which 
electrons get liberated from the atoms and the 
metallic conductivity is beginning to set in 
\cite{ex1}. Given the importance of the subject 
one may hope  the conditions necessary for 
reaching densities above $100~g/cm^3$ 
could be achieved in not too far future.
(iii) The deuteron matter with  $\rho\sim 10^3~g/cm^3$
is expected to  be present in localized regions within low 
temperature ($T\lsim  10^5~K$) low-mass brown dwarfs.
These are compact astrophysical objects 
that have not  managed -- because of their low mass and temperature -- 
to ignite nuclear fusion (i.e., they are ``failed stars'').  
As we will argue, these low-mass, cold brown dwarfs should  contain 
the deuteron quantum liquid in a thin layer in their interiors.
(iv) Theoretically,  the deuteron quantum liquid 
seems to be a richer system than its   helium-4 counterpart. 
Due to deuterons being spin-1, the spectrum of small fluctuations of 
the deuteron liquid exhibits properties similar to those of 
antiferromagnetic superconductors, as it will be shown in the paper.
Moreover, the deuteron condensate represents an amusing example for  
which a consistency of counting of the Nambu-Goldstone  modes in 
non-Lorentz-invariant theories \cite{Nielsen} can be tested.
(v) Our formalism can automatically be applied to condensates of 
other charged spin-1 particles, or  to neutral spin-1 particles 
with contact,  and/or   spin-spin, and/or magnetic dipole  interactions, 
by taking various limits.

\vspace{0.1cm}  

Before we turn to the field theory description in the next section, it is instructive
to summarize the conditions for the condensation to take place.

At sufficiently high temperatures (but still significantly below the temperature 
of efficient thermo- and pycno-nuclear fusion), 
say at $T\lsim 10^5~K$,  the system of deuterons 
would form a classical Bose gas, even though the electrons may already be  in a 
quantum degenerate state. At lower temperatures,  however, transitions of deuterons
to  different states is possible --  they may crystallize, 
or alternatively, may form a quantum liquid.

Classical crystallization occurs at  temperatures for which the Coulomb 
interaction energy of a pair of nuclei 
is about 180 times greater than the classical thermal energy of 
two degrees of freedom  \cite{cr1}. This leads to  
crystallization  temperature for deuterons 
(see, \cite{gp} for recent discussions):
\beq
T_{\text{cryst}}\simeq 1000 \cdot \rho^{1/3}~K, 
\label{tempCryst}
\eeq
where density $\rho$ is in units of $g/cm^3$.
At high-enough  densities, however, before crystallization temperature is reached 
from above,  the deuterons may undergo transition into the charged 
condensate. This transition would take place  around certain  temperature $T_c$ for  
which the thermal  de Broglie wavelength of the nuclei becomes greater than
an average inter-particle separation between them. 
If $T_c$ significantly exceeds the  crystallization temperature, 
then, before the system gets a chance to crystallize,  
the de Broglie wavelengths of the ions would begin to overlap
leading to quantum-mechanical indistinguishability of the nuclei.  
If so, the standard arguments (see, e.g., \cite {Leggett}) 
suggest that the bosons would prefer to occupy one and the same quantum 
state, i.e., would prefer to condense. 
An  estimate  \cite {gr1,gp}
for the condensation temperature is 
\beq
T_c \simeq 350 \cdot \rho^{2/3}~K.
\label{tempCond}
\eeq
The condensation temperature grows faster with $\rho$ than
the crystallization temperature does\footnote{For a discussion of 
uncertainties in  (\ref{tempCryst}) and (\ref{tempCond}), 
which should not be affecting much our conclusions, 
see \cite{ggshifman,grrev}.}. In particular, comparing (\ref{tempCryst}) 
and (\ref{tempCond}), we conclude that for densities greater than 
$100~g/cm^3$, temperature $T_c$ significantly exceeds $T_{\text{cryst}}$, 
and hence deuteron condensation is favored  
upon gradual cooling of the system\footnote{At yet lower temperatures, when the 
crystalline state also becomes available, there may exist spontaneous transition 
from the condensate to the crystalline state. However, this process is  strongly  
suppressed \cite {ggshifman}.}. 

\vskip 0.3cm

In the next section, based on  universal symmetry principles,  
we formulate an order-parameter effective field theory 
describing the deuteron condensate. We 
discuss the symmetry breaking patterns by the condensate, and count the 
broken generators. In section 3 we obtain the solution 
that describes a spin-0 state of condensed spin-1 deuterons,  
and discuss small fluctuations 
about  that solution.   Interestingly, the  fluctuations contain a massive
mode,  which is well below the ultraviolet (UV) cutoff of the effective theory,
as well as two massless modes with linear dispersions.  
The massive mode is a reflection of the 
Meissner effect, while the massless modes are similar to antiferromagnetic 
spin-waves. Hence, the condensate  has  properties similar to a
antiferromagnetic superconductor, but has one order parameter. 
Moreover, the massless modes are Nambu-Goldstone (NG) bosons of 
spontaneously broken rotational symmetry.
We discuss how  counting of the NG  modes agrees with the rules 
of Goldstone's theorem applied to  non-Lorentz-invariant systems.
In section  4 we study thermodynamic properties of a deuteron quantum 
liquid and briefly discuss its applications to low-mass and low-temperature brown dwarf 
interiors. Three  appendices summarize various important technical  details 
which, if placed in the main text, would overwhelm the presentation.

\vskip 1cm 
\begin{center}
{\large \bf 2. Effective Field Theory and Symmetry  Breaking} 
\end{center}
\vskip 0.2cm

The condensed state of deuterons discussed in the previous section 
will be described here by  a complex vector order parameter  
$W_i({\bf r}, t), ~i=1,2,3$,
that can slowly vary at scales much greater than the average inter-deuteron  
separation, $d\sim J_0^{-1/3}$ (hereafter $J_0$ denotes the deuteron 
average  number density, $J_0\sim 10^{-6}~MeV^3$).  A  non-zero background  value of the order 
parameter --its expectation value in the state with a macroscopic occupation  
$\langle W_i({\bf r}, t)\rangle $ --  is a signature of the 
condensate. Small perturbations about the background, 
$\delta W_i \equiv  W_i- \langle W_i\rangle $, describe long-wavelength  
collective excitations of the system.

In this section  we will set up  an  effective Lagrangian for the order 
parameter 
$W_i({\bf r}, t)$. Since the deuteron mass, $m_H\sim 1.876 ~{\rm GeV}$,  is much 
greater than any energy/momentum  scale that we'll be dealing  with,
the non-relativistic approximation in $m_H$ will be adopted 
(see Appendix A for the relativistic Lagrangian). The short-distance cutoff
of the effective Lagrangian is set by the scale of interparticle separation
$d\sim J_0^{-1/3}$.  As we will see, the effective Lagrangian 
will capture mass-less,  as well as massive collective excitations 
of the system. The latter  have  masses $\sim g (J_0/m_H)^{1/2}$,
where $g$ denotes the gauge coupling constant.  Because of the hierarchy 
$J^{1/3}_0\ll m_H$,  the above  mass scale is 
parametrically  supressed in comparison with the UV cutoff of 
the theory,   $d^{-1}\sim J_0^{1/3}$, and hence, can be calculated 
reliably within the effective Lagrangian approach.

The effective  Lagrangian has to be invariant w.r.t. all the internal and 
space-time symmetries appropriate for the physical system at hand. 
In particular, it should possess the local $U(1)_{EM}$ 
symmetry of electromagnetism, the global subgroup of which guarantees 
the electric charge 
conservation. It should also posses a global $U(1)_{s}$ symmetry w.r.t. 
$W_i({\bf r}, t) \to W_i({\bf r}, t)e^{i\beta} $ with an arbitrary 
constant $\beta$, as the latter guarantees conservation of the  number of deuterons. 
The above two global symmetries,  once put together,  imply that the electron 
number is also conserved.  As to  space-time symmetries, the Lagrangian     
should be invariant w.r.t. spatial rotations and reflections  
represented by the $O(3)$ group; it  should also be  Galilean invariant.   
Finally, in the limit when all the relevant  coupling 
constants are set to zero, the 
equations of motion should reduce to the ordinary Schr\"odinger equation for 
the order parameter $W_i$.

The Lagrangian  that satisfies all the above conditions
and describes  the bosonic sector of the theory
reads as follows [we use the ``mostly-minus'' metric convention]:  
\beq
\mathcal{L}_1=-\frac{1}{4}F_{\mu \nu}F^{\mu \nu}-\frac{i}{2}(W^{*i}D_0 W_i-
W^i(D_0 W_i)^*)-
\frac{1}{2 m_H}|D^i W_j|^2-\nonumber \\ -\frac{\lambda_q}{8m_H^2}
(W^{*i}W_i)^2-\frac{\lambda_s}{8m_H^2}(\epsilon^{ijk}W^*_jW_k)^2-
ig\frac{(1-\kappa)}{2m_H} W^*_{i} W_{j}F^{ij}\,.
\label{nrl1}
\eeq
The covariant derivative is defined as $D_\mu=\partial_\mu-igA_\mu$, $g$ 
is a charge of $W$.  The  effective mass of a deuteron
in the interacting system is denoted by $m_H$.  We will assume 
that this mass --
being the highest scale -- does not get renormalized too much  
by interactions and set 
$m_H\sim 2~{\rm GeV}$. The parameters 
$\kappa$,   $\lambda_s$, $\lambda_q$  are dimensionless 
coupling constants. The last term in the Lagrangian parametrizes 
the effects  of the magnetic moment of charged deuteron nuclei 
for which  we included the tree-level as well as loop-generated terms  
(the latter is  parametrized by $\kappa$).  Furthermore, 
the $\lambda_s$- term describes the spin-spin interaction, 
while the  $\lambda_q$- term parametrizes the contact 
repulsion of the deuterons\footnote{Note that in a ``fundamental theory''
the $\kappa$,   $\lambda_s$, $\lambda_q$ -terms  need not be included in 
the tree-level Lagrangian as they'll be generated via loop corrections. 
In the effective Lagrangian, since we're not going to be calculating the 
loops, these terms were  introduced  with arbitrary 
coefficients in (\ref {nrl1}).}.

We have included in (\ref {nrl1}) only leading order terms. 
There are higher dimensional terms 
in the effective Lagrangian that are consistent  with the symmetries imposed.  
A  systematic way to include those terms is
to turn to the canonically normalized fields 
(replace  $W\rightarrow W\sqrt{2m_H}$), and 
then write down the  higher dimensional operators 
suppressed by appropriate powers of the short distance (UV) 
cutoff,  $d^{-1}\sim J_0^{1/3}$.  
Because of the hierarchy $J_0^{1/3}\ll m_H$, the higher dimensional terms will 
not play a significant role  for the low energy 
excitations of the system and will be ignored.

In addition to (\ref {nrl1}) the full theory also includes 
terms describing electrons of  some effective 
mass $m_e$, for which we'll be using the Dirac Lagrangian
\beq
\mathcal{L}_2=i\bar\psi \gamma^{\mu}(\partial_\mu+igA_\mu)
\psi-m_e\bar\psi\psi.
\label{Dirac}
\eeq
The total Lagrangian reads as follows:
$
%\beq
\mathcal{L}^{NR}=\mathcal{L}_1+\mathcal{L}_2.
%\eeq
$
For the reasons discussed above, we will also ignore  higher 
dimensional terms which couple electrons directly 
to the order parameter (an example being  
$W^*W \bar\psi\psi$). Even though we retain the Dirac 
Lagrangian in the canonical form (\ref{Dirac}), 
the electrons themselves are heavy and decoupled from the low-energy 
dynamics. What is not decoupled are the gapless excitations of the electrons  
near the Fermi surface.  To begin with,   we treat the fermionic 
fluctuations in the 
Thomas-Fermi (TF) approximation, although this will be refined 
later on to include a loop correction  due to those near-the-Fermi-surface 
gapless modes. In the TF  approximation we average the  fermionic 
four-current 
as follows: $J^\mu\equiv\langle\bar\psi\gamma^\mu\psi\rangle=J_0\delta^\mu_0$.
This sets up a nonzero electron number-density $J_0$ 
in the ground state. Note that this state  
does not break the global  $U(1)$ symmetry responsible for 
the fermion number conservation. 

As we will show in the next section  the  solution
\beq
\langle A_\mu\rangle=\frac{\lambda_q v^2}{2m_H}\delta^0_\mu, 
\qquad \langle W_{\mu}\rangle =(0,0,0,v),
\label{vac10}
\eeq 
where $v\equiv\sqrt{J_0/2m_H}$, represents a consistent 
classical background of the theory. It describes a 
state of condensed nuclei with   zero net spin, in which the 
individual deuteron spins are  aligned  or anti-aligned with  
the preferred  direction set by a vector ${\bf v} = (0,0,v)$.
Such a spin  alignment,  and the resulting spin-waves 
discussed in the next section,  are similar to those in  
antiferromagnetic materials. On the other hand, the properties of 
the electromagnetic field  in this substance 
will be shown to be similar to those in superconductors. 

Note that a nonzero expectation value of $A_0$ in (\ref {vac10}) 
shifts the value of Fermi energy, however, this correction is typically 
of the order $\sim \lambda_q 10^{-12}~MeV$, and will be ignored.  

Before turning to the next section,  let us discuss the pattern of 
symmetry breaking by the background (\ref{vac10}) in  
a few specific cases:

\vspace{0.1cm}

{\it Case ``a": $\lambda_q \neq0,\lambda_s\neq0, \kappa\neq 1$ }. 
The internal part of the symmetry group $G$ of the theory 
is given by $G_{int}=U(1)$, whereas the 
relevant-to-us part of the space-time symmetry consists of the group of reflections 
and  spatial rotations $G_{st}=O(3)_{s}$ (the Galilean and space-time translation 
subgroups of $G$ are left unbroken by the condensate, and are ignored 
in the further discussions). Upon vector condensation (\ref{vac10}), 
the symmetry group is broken down to spatial $O(2)_s$, 
corresponding to rotations 
and reflections in the plane  perpendicular to $\mathbf{v}=(0,0,v)$. 
Therefore,  $U(1)\times O(3)_{s} \rightarrow O(2)_s$, and the 
number of spontaneously broken continuous symmetry generators 
ends up being equal to 3.

\vspace{0.1cm}

{\it Case ``b": $\lambda_q \neq0,\lambda_s\neq0, \kappa = 1$ }. 
The internal symmetry group gets enhanced by an additional factor 
of $O(3)_i$ corresponding to a simultaneous internal rotation of $W_i$ 
and $W^*_i$ by orthogonal three-by-three matrices. The diagonal part 
of  $O(3)_{i}\times O(3)_s$ gets broken by the vacuum (\ref {vac10})
to the diagonal $O(2)_d$,  whereas the  orthogonal combination 
$O(3)_a$ remains unbroken.  This leads to the following pattern of 
symmetry breaking  $ U(1)\times O(3)_{i}\times O(3)_s 
\rightarrow O(3)_a \times O(2)_d$, and consequently 
to three broken generators.

\vspace{0.1cm}

{\it Case ``c": $\lambda_q \neq0,\lambda_s = 0, \kappa = 1$ }.
Internal symmetry is enhanced further to $G_{int}=O(6)$. 
The symmetry breaking  pattern is then  $O(6) \times 
O(3)_s \rightarrow O(5) \times O(3)_{lin}$,
where  $O(3)_{lin}$ is a linear combination of a
subgroup of $O(6)$ and  $O(3)_s$. This results  
in five broken generators.

\vspace{0.1cm}

The Lagrangian (\ref{nrl1}) describes a real vector field $A_\mu$ 
with two dynamical components,  and a complex massive vector field $W_i$ 
with six components three of which are heavy degrees of freedom 
(mass $\propto m_H$)  which  should be ignored; 
this  leaves  altogether five potentially light degrees of 
freedom. Irrespective of the Case ``a'', ``b'' or ``c'' described above,
the $U(1)$ electromagnetic symmetry is spontaneously broken by the condensate, 
hence, the longitudinal NG boson will become massive to 
serve as a helicity-0 component of a massive photon. 
Therefore,  two NG bosons should remain and  satisfy 
the counting rules for the different symmetry breaking patterns 
given in Cases  ``a'', ``b'' or ``c''. At a first sight this seems unlikely, 
however, as we will see in the next section, that the dynamically obtained 
NG  modes will obey the rules of counting   
of the Goldstone theorem  for Lortentz-non-invariant  
models \cite {Nielsen}; these rules are quite different  
from those in relativistic theories.

Last but not least, in this and the next section  we're 
discussing  the effective Lagrangian at zero temperature  (finite temperature effects in charged scalar condensate have been recently studied in \cite{dolgov}), 
even though what we actually have in mind is  temperature  
much lower than  the deuteron condensation temperature,  
but higher than the room temperature. 
A caution has to be exercised in this regard  -- 
at a certain temperature $T_e\ll 10^{-6}~K$ (see the next section) 
there will be additional phase transition in which the Cooper 
pairs of near-the-Fermi-surface 
electrons will form via the  Kohn-Luttinger  
mechanism \cite {Kohn} (see, \cite {grrev} for a brief discussion of 
this effect in the presence of the BE condensate). 
Then, below $T_e$  a new NG boson will 
emerge  as a result of spontaneously 
broken fermion-number symmetry triggered 
by the  electron-electron order parameter. 
In conventional  superconductivity  an analogous  
NG Boson combines with a photon to form a massive photon.  
In the present case, however, this  
role is already played by the longitudinal mode 
associated with spontaneously broken $U(1)_s$ 
boson-number symmetry discussed above\footnote{In reality one should 
consider linear combinations of these two NG bosons; 
one of the combinations  will become massive and another one will 
remain massless.  However, since $gJ_0/m_H  \gg  T_e^2$, the mixing 
effects are negligible.}.
On the other hand, since  $T_e$ is  close to zero 
for any system discussed in this work  we'd expect temperature to be 
way above $T_e$. Therefore,  in our discussions 
we will not invoke  the Cooper pairing  mechanism with the 
fermion-number breaking condensate, and will ignore the finite 
temperature effects.

\vskip 1cm 
\begin{center}
{\large \bf 3. Background and its Small Fluctuations}
\end{center}
\vskip 0.3cm

Below we derive the background solution  and study fluctuations about it.
It is convenient to use a slightly different form of the Lagrangian 
(\ref{nrl1}) in which we ``integrate in'' an auxiliary field $W_0$, 
and also perform the following rescalings of the fields,  
$W\rightarrow W\sqrt{2m_H}$, $A\rightarrow A/g$. Then, 
the part of the Lagrangian relevant for studying the bosonic 
background and its fluctuations reads as follows:
\beq
\mathcal{L}_1=-\frac{1}{4g^2}F_{\mu \nu}F^{\mu \nu}-\frac{1}{2} \Sigma^*_{ij}
\Sigma^{ij}+im_HW^{*i}\Sigma_{i0}-im_HW^{i}\Sigma^*_{i0}  
+m^2_HW^*_0 W^0-\nonumber 
\\ -J_{\mu} A^{\mu}-\frac{\lambda_q}{2}(W^{*i}W_i)^2-\frac{\lambda_s}{2}
(\epsilon^{ijk}W^*_jW_k)^2+i\kappa W^{*i}W^jF_{ij}, \qquad
\label{nrl'}
\eeq
where $\Sigma_{ij}=D_i W_j-D_j W_i$ denotes the gauged  field strength of $W$,
and the fermions have been treated in the mean-field approximation, 
as mentioned before\footnote{A derivation of  (\ref{nrl'})  as 
a non-relativistic limit of a relativistic theory in 
given in Appendix B.}. There are no time derivatives of the 
$W_0$ field  in (\ref {nrl'}), and as noted before, 
it is an auxiliary variable which can be integrated out to get back 
the Lagrangian (\ref {nrl1}).

Equations of motion for the fields $W^*_0,W^*_i,A_0,A_i$ are 
straightforward 
to obtain from (\ref{nrl'}). They read  respectively as follows:
\begin{eqnarray}
im_HD^iW_i+m^2_HW_0=0, \nonumber
\\
D^j\Sigma_{ji}-im_HD_0W_i+im_H\Sigma_{i0}-(\lambda_q+\lambda_s)
(W^{*j}W_j)W_i+\lambda_s(W^jW_j)W^*_i+\nonumber 
\\ i\kappa W^jF_{ij}=0, \nonumber
\\ 
\frac{1}{g^2}\partial ^j F_{j0}-2m_HW^{*i}W_i=J_0,\nonumber
\\
\frac{1}{g^2}\partial ^\mu F_{\mu i}+iW^{*j}\Sigma_{ji}-iW^j
\Sigma ^*_{ji}+m_H(W^*_iW_0+W_iW^*_0)-i\kappa\partial^j(W^*_jW_i-W^*_iW_j)=0.
\nonumber
\end{eqnarray}
It is not difficult to check that these equations have the following 
classical solution 
\beq
A_\mu-\partial_{\mu}\alpha=\frac{\lambda_q v^2}{2m_H}\delta^0_\mu, 
\qquad W_{\mu}=(0,e^{i\alpha}\mathbf{v}),
\label{givi}
\eeq
where  $\mathbf{v}$ is an arbitrarily-directed real vector with the norm 
$v\equiv\sqrt{J_0/2m_H}$,  and $\alpha$ is an overall phase of the 
vector field $W_\mu$. As discussed in the previous section,  
we choose  our background solution  to be:
\beq
\langle A_\mu\rangle=\frac{\lambda v^2}{2m_H}\delta^0_\mu, 
\qquad \langle W_{\mu}\rangle =(0,0,0,v).
\label{vac1}
\eeq 
This solution describes a state of a large occupation number 
of bosons, the net electric charge of which is compensated
by the electron charge density $J_0$. The quantities that acquired the vacuum expectation values are gauge-invariant, as can be seen from (\ref{givi}). The gauge-choice given by (\ref{vac1}), on the other hand, corresponds to the unitary gauge.
 
To study the spectrum of small perturbations, 
it is useful to work in terms of the gauge-invariant variables
\beq
A_\mu-\partial_\mu\alpha _3,\qquad |W_3|, \qquad W_{0,1,2} ~e^{-i\alpha_3},
\label{gf}
\eeq
where $\alpha_3$ denotes the phase of $W_3$ (the same results would have 
been obtained by choosing the unitary gauge).  As mentioned before, 
we treat the fermionic fluctuations in the Thomas-Fermi 
approximation:  This can be accounted  
for in the equations of motion by increasing the  
coefficient of the $A_0$ perturbation 
by $m_0^2/g^2\equiv(3\pi^2 J_0)^{2/3}/\pi^2$ in the  
ultra-relativistic  regime for the electron Fermi-gas,
or by   $m^2_0/g^2\equiv m_e (3\pi^2J_0)^{1/3}/\pi^2$ 
in the non-relativistic regime.

The equations of motion for the excitations on the condensate, 
as well as the procedure of obtaining the linear spectrum 
are briefly summarized in Appendix C. The calculations are straightforward 
but very tedious. We find that the dispersion relations for the relevant, 
low-energy modes are given by the following:

\textit{Longitudinal polarization  of the massive photon.} After symmetry 
breaking, one combination of the fields becomes the longitudinal 
mode of the massive photon, with the dispersion relation given by
\beq
\omega_1^2 \simeq 2 v^2 g^2 \left(1+\frac{\mathbf{k}^4+2 
\lambda_q  v^2 \mathbf{k}^2 }{8 g^2 m_H^2 v^2}\right).
\label{bdisp}
\eeq
An identification of the corresponding mode is done  
by invoking the following arguments:  first, we see that 
this dispersion is independent of the spin-spin interaction 
constant $\lambda_s$,  as well as the magnetic moment interaction  
constant ($1-\kappa$), suggesting that this is a longitudinal mode. 
Second, in the limit of zero electromagnetic coupling, $g\to 0$, 
the dispersion relation for small ${\bf k} $ is linear in  
$ {\bf k} $  and is determined by the ``contact repulsion'' 
constant  $\lambda_q$, as it should be for a longitudinal mode
\footnote{In the denominator of the second term in  
(\ref {bdisp})  we ignored  terms proportional to $m_0^2 {\bf k}^2$ and
$\lambda_qv^2 m_0^2$. These terms are always sub-dominant to the 
term retained  in the denominator.}. Third, if in addition we take 
the limit  $\lambda_q \rightarrow 0$ we obtain the dispersion relation for a 
small fluctuation above a Bose-Einstein (BE)  
condensate of free  particles, $\omega_1 \simeq \mathbf{k}^2/2m_H$.
 
\textit{Transverse polarizations of the photon}. A complete set of 
dispersion relations for the two transverse modes of the photon are 
given in Appendix C by $\omega_{2,3}$ in (\ref{zust!}). 
In the limit when  all the coupling constants are  switched off
we obtain ordinary photon dispersions.  In the gauge theory 
the transverse modes acquire mass gaps, which can easily be seen by 
setting $\k \rightarrow 0$. Expanding $\omega_{2,3}$ in (\ref{zust!}) 
in powers of the momentum for which $|\k|\ll gv$, one obtains 
\begin{equation}
\omega_{2}^2\simeq 2g^2v^2+\mathbf{k}^2+ 
\frac{(1-\kappa)^2}{4 m^2_H}\mathbf{k}^4 ,
\label{tr1}
\end{equation}
\begin{equation}
\omega_{3}^2\simeq 2g^2v^2+\mathbf{k}^2+ 
\frac{(1-\kappa)^2}{4 m^2_H}\frac{(\k\cdot\mathbf{v})^2}{v^2}\mathbf{k}^2.
\label{tr2}
\end{equation}
Hence, the transverse modes are massive with the value of the mass
coinciding with that of the longitudinal mode (\ref {bdisp}).
A seeming independence of the  last  terms in the 
dispersion relations (\ref{tr1}), (\ref{tr2}) from 
the gauge coupling constant is due to the specific nature of 
the expansion $|\k|\ll gv$ -- the dependence on $1/gv$ shows up in 
higher order terms. The exact expressions (\ref{zust!}) 
do no exhibit any pathological behavior. 

In  the opposite limit $|\k|\gg gv$, one may 
perform the $1/m_H$ expansion, leading to the following 
dispersions
\begin{equation}
\omega_{2}^2\simeq 2g^2v^2+
\mathbf{k}^2+2g^2v^2\frac{(1-\kappa)^2}{4 m^2_H}\mathbf{k}^2,
\end{equation}
\begin{equation}
\omega_{3}^2\simeq 2g^2v^2+\mathbf{k}^2+
2g^2v^2\frac{(1-\kappa)^2}{4 m^2_H}{ (\mathbf{k}\cdot\mathbf{v})^2\over v^2} .
\end{equation}
As we see, propagation of the transverse modes in general 
is affected by the magnetic moment interaction. However, 
$\omega_{3}$ depends on the preferred 
direction of the spin-antispin alignment set by 
the vector ${\bf v}$, while  $\omega_{2}$ 
does not. As a result, for momenta perpendicular to the preferred direction,
frequency $\omega_{3}$  is independent of $(1-\kappa)$. 
On the other hand, when the direction of propagation coincides 
with the  preferred direction,  the expressions for 
$\omega_{2}$ and $\omega_{3}$ coincide too; this is   
a manifestation  of the remaining rotational symmetry 
in the plane perpendicular to this direction.

\textit{NG bosons of broken rotational symmetries.} The two massless 
NG modes constitute the rest of the spectrum, their dispersion relations 
being given by $\omega_{4,5}$ in (\ref{zust!}).  In the zero coupling 
limit these reduce to  the dispersion relations of small fluctuations over 
the BE condensate of free bosons, $\omega_{4,5}\simeq k^2/2m_H$. 
In the interacting theory we perform the expansions similar 
to the ones used for the transverse modes,  and obtain
\begin{equation}
\omega^2_4\simeq \frac{\mathbf{k}^4}{4m^2_H}+
\lambda_s \frac{\mathbf{k}^2 v^2}{2 m^2_H}- 
\frac{(1-\kappa)^2}{4 m^2_H}\mathbf{k}^4 ,
\label{inst1}
\end{equation}
\begin{equation}
\omega^2_5\simeq \frac{\mathbf{k}^4}{4m^2_H}+
\lambda_s \frac{\mathbf{k}^2 v^2}{2 m^2_H}-
\frac{(1-\kappa)^2}{4 m^2_H}\frac{(\k\cdot\mathbf{v})^2}{v^2}\mathbf{k}^2,
\label{inst2}
\end{equation}
in the $|\k|\ll gv$ limit, while for the   $|\k|\gg gv$ case we obtain 
\begin{equation}
\omega^2_4\simeq \frac{\mathbf{k}^4}{4m^2_H}+
\lambda_s \frac{\mathbf{k}^2 v^2}{2 m^2_H}-
2 g^2v^2\frac{(1-\kappa)^2}{4 m^2_H}\mathbf{k}^2 ,
\label{inst3}
\end{equation}
\begin{equation}
\omega^2_5\simeq \frac{\mathbf{k}^4}{4m^2_H}+\lambda_s 
\frac{\mathbf{k}^2 v^2}{2 m^2_H}-2 g^2v^2 \frac{(1-\kappa)^2}
{4 m^2_H}{(\mathbf{k}\cdot\mathbf{v})^2\over v^2}.
\label{inst4}
\end{equation}
A few comments are in order here: 
(i) As in the case of the transverse modes, seeming independence 
of the leading terms in NG dispersion relations (\ref{inst1}), (\ref{inst2}) 
from the gauge coupling constant is an artifact of the expansion in 
$|\k|/gv\ll 1$. The spectrum is 
stable for $\kappa\in[0,2]$ and $\lambda_s\geq0$. 
(ii) The $\lambda_s$-dependent  quadratic terms in the NG 
dispersion relations are  due to the spin-spin interaction term ($\lambda_s$-term)
in the original Lagrangian (\ref {nrl'}). (iii) The last, $(1-\kappa)$-dependent 
terms in (\ref{inst3}) and (\ref{inst4}),
are due to magnetic moment interactions of 
oppositely aligned spins of charged 
deuterons. Since these interactions are attractive, 
the minus sign in front of the last terms in 
(\ref{inst3}) and (\ref{inst4}) appears. In the exact expressions 
(C-VII) these negative terms can never overcome the positive ones, 
and the spectrum is stable. 
(iv) The  dependence  on the gauge coupling constant seems to 
disappear  from the dispersions of the NG modes for $\kappa=1$.
However, in the system of deuterons and electrons 
the effective constant of the spin-spin interactions,  
$\lambda_s$, and the anomalous magnetic moment of deuteron, $\kappa$,  
will be proportional to positive powers of 
the gauge coupling. (v) As in the  case of the transverse modes, 
for momenta aligned with the preferred direction, 
the dispersion relations for the two NG bosons 
coincide,  as a reflection of the remaining $O(2)$ 
symmetry group.

All the described properties above suggest that the obtained gapless
states are NG bosons of spontaneously broken rotational symmetry.
Their properties are very similar to spin-waves in antiferromagnets; 
in the present case they emerge due to spins of the charged nuclei.

It is time now  to check a consistency of 
counting of the NG bosons with  Goldstone's
theorem for Lorentz  non-invariant systems. This counting is 
rather different  from that in relativistically invariant  
theories, and goes as follows \cite{Nielsen}: 
if one denotes the number of broken generators of continuous 
global symmetry by $q$, then the number of NG bosons should 
satisfy the following inequality 
\beq
q\leq n_1+2n_{2}, 
\label{ineq}
\eeq
where $n_1$  is the numbers of NG modes 
with frequencies  proportional to an odd 
power of momentum  in the long wavelength limit, 
(e.g., $\omega \propto |{\bf k}|$; these are called "type I"), 
while $n_{2}$ is the  numbers of NG modes  with frequencies  
proportional to an even power of momentum in that  limit 
(e.g., $\omega \propto |{\bf k}|^2$; called "type II").

One can easily check, that (\ref{ineq}) is remarkably 
satisfied by the NG modes found for  the cases 
``a'', ``b'', and ``c''  discussed at the end of 
the previous section. In the Case ``a'' or  ``b'',  
there are three  broken  generators one of which is Higgsed, hence, $q=2$.
This is matched by  two type I NG modes, which saturate 
the  inequality (\ref {ineq}). 
If we impose a further  condition that $\lambda_s=0$ -- 
as in the Case ``c''--  then we end up with 5 broken generators 
one of which is Higgsed, leading to $q=4$.  
On the other hand, both NG bosons found above end up being  of 
type II in this limit, hence the inequality (\ref {ineq}) is saturated again.

One more interesting example is that of Case ``a'' with an additional condition 
$\lambda_s=0$. The symmetry breaking pattern in this case is identical to 
that of Case ``a''. However, the dispersion relations for 
the NG modes, (\ref {inst1})   and (\ref {inst2}), become of type II, 
i.e,  we get $n_1 =0$ and $n_2 =2$. Clearly,  in this case the inequality 
(\ref {ineq}) is not saturated,  but is satisfied.

As  discussed at the end of the previous section, there will 
be one more longitudinal NG mode  that emerges at extremely 
low temperatures (very close to the absolute zero) 
as a result of  spontaneous breaking of the femion-number symmetry 
due to Cooper pairing of the near-the-Fermi-surface 
electrons via the Kohn-Luttinger mechanism.

\vspace{0.1in}

We end this section by considering more refined effects of the near-the-Fermi-surface 
electrons.  So far  we have taken these effects into account in the Thomas-Fermi 
approximation.  This approximation, however, fails  to capture one of the most important 
properties of the  fermion system,  namely that the  gapless modes near the 
Fermi surface can give rise to {\it long range}  interactions for charged particles 
via the loop effects. 

One of the manifestations  of this  is the well-known Friedel potential \cite {Walecka},
and the Kohn-Luttinger effect itself. It is important to see how these considerations--
traditionally done for a system with  fermions only  -- 
are affected by the presence of the spin-1 charge condensate 
(for spin-0 see \cite {gr1}). 

To study these effects we consider a potential between two static charges (e.g., 
two impurities,  such as helium-3 or helium-4 nuclei).
The zero-zero component of the frequency-independent momentum-space Green's 
function in our case  reads as follows (for simplicity, we take $\lambda_q=0$):
 \beq
G(\omega =0,\mathbf{k},\lambda=0)=
\frac{\mathbf{k}^2}{\mathbf{k}^4 + 4M^4},
\eeq  
where $M\equiv(2g^2m^2_Hv^2)^{1/4}$.  In the coordinate space this would give 
rise to an exponentially suppressed oscillating potential $(e^{-Mr}/r)
{\rm cos} (Mr)$.  We now include the contribution  of  the electron via one 
loop vacuum polarization diagram. This leads to two effects \cite {Walecka}: 
(i) the pole of the Green's function  gets shifted (this is what leads to 
the Debye screening, and is the effect captured in the Thomas-Fermi approximation 
by the parameter $m_0^2$); (ii)  the gapless excitations near the Fermi surface, 
as in the case of the Friedel potential, lead to branch-cuts in the complex 
$|{\bf k}|$-plane.  These branch-cuts give rise  
to the {\it long range} piece in the position-space 
potential,  which in the case with relativistic fermions (taken
for simplicity), ends up  being  similar to the one obtained in \cite{gr1}
\beq
V_{stat}= \alpha_{\rm em}  {Q_1Q_2} \left (  { e^{-Mr}\over \,r}
{\rm cos} (Mr)\, + {4 \alpha_{\rm em} 
\over \pi} { k_F^5{\rm sin}(2k_Fr)\over M^8r^4}\right )\,. 
\label{potential}
\eeq
Here $Q_{1,2}$ denote the charges of static sources (impurities),  
$\alpha_{em}$ is the fine structure constant and $k_F$- Fermi 
momentum of the system. 

As in  the case of the charged spin-0 condensate,  the obtained long-range potential 
is suppressed in comparison with the standard Friedel potential. The reason for this
suppression  is the subtraction due to an exchange of 
the longitudinal polarization of the photon, as 
discussed in detail in \cite{gr1}.

The potential (\ref {potential}) is sign-indefinite and one may wonder whether its 
attractive part could trigger some instability near the Fermi surface to 
form Cooper pairs.
In fact, the analogous oscillatory behavior in the  interaction potential between 
electrons is what gives rise to the Kohn-Luttinger effect. However, as we see
the potential is too small  --  a ``typical'' value for it is 
 $10^{-16}~ {\rm MeV} \sim 10^{-6}~K$, and therefore  any tiny finite temperature 
will wash-out the effects of this potential.  This small  value of the potential 
is also the reason for the temperature of the Kohn-Luttinger type 
transition to be  very  low -- it should be that $T_e\ll 10^{-6}~K$.

\vspace{1cm}
\begin{center}
{\large \bf 4. Some Properties of the Condensate}
\end{center}
\vspace {0.3cm}

Already at densities $\rho\sim 10^3~g/cm^3$ (which is well above atomic 
densities for deuterium), the temperature of condensation significantly 
exceeds that of crystallization (the critical density, at which 
$T_c \simeq T_\text{cryst}$ is about an order of magnitude lower).
Hence, according to the arguments given in section 1, at these densities 
the charged deuteron condensate is expected to form at 
$T< T_c\sim 3.5\cdot 10^4~K$.

In order to study thermodynamics of this substance at $T \ll T_c$, 
we need an expression for the temperature dependence of its 
specific heat (\textit{per ion}). The latter 
will be determined by the massless NG modes\footnote{The gapped modes -- two 
transverse and one longitudinal--  are effectively "decoupled" due to their mass 
$\sqrt{2}gv\sim 5\cdot 10^4 K$ being greater than the condensation temperature, while fermion
contribution to the specific heat is negligible.}. 
The expression for the specific heat (in the units of the Boltzmann constant) is readily obtained from 
\beq
c_v=\frac{1}{(2\pi)^3J_0}\int^{k_c} d^3k 
\frac{\omega^2 e^{\omega/T}}{(e^{\omega/T}-1)^2T^2}.
\label{sh}
\eeq
Here $k_c\equiv 2\pi/d$ represents the natural momentum cutoff in 
the effective theory. For densities mentioned above, $k_c\sim 10^{-1}~MeV$. 
At the momenta close to the cutoff, the dominant contribution in  the 
NG dispersion  relation is given by  the first terms in (\ref{inst3})-(\ref{inst4}), 
which are of the order $\omega_{4,5}\sim 10^{-6}~MeV\simeq 10^{4}~K$.

We note that for the momenta greater than  $k_0 = \sqrt{\lambda_s}\cdot 10^{-5}~MeV$, 
the $k^2/2m_H$ term  still dominates in the NG dispersion relations for 
most of the integration range in (\ref{sh}).  A simple numerical analysis shows, 
that the contribution from the momentum domain  $k\in(0,k_0)$  is negligible  as long as 
$T\gg \lambda_s v^2/m_H \sim \lambda_s\cdot 10^{-3}~K$, which is assumed to be 
the case here.

Therefore, the thermodynamics of the NG gas is determined by the free particle 
dispersion in the temperature range at hand,   $ \lambda_s\cdot 10^{-3}~K< T< 10^4~K$. 
Using these estimates, we derive the expression for the 
specific heat per ion 
\beq
c_v=\frac{1}{J_0}\left(\frac{2m_H T}{\pi}\right)^{\frac{3}{2}}\frac{15~ \zeta(5/2)}{32}.
\eeq
At temperatures $T\ll \lambda_s v^2/m_H\simeq \lambda_s\cdot 10^{-3}~K$,  
on the other hand, $c_v$ is determined by the $\lambda_s-$ term in the NG 
dispersion relations. Specific heat per deuteron in the ultra-low temperature 
limit therefore, is given by
\beq
c_v=\frac{16\pi^2}{15\lambda_s^{3/2}}\frac{m^{9/2}T^3}{J_0^{5/2}}.
\eeq  
This is true as long as temperature is above the point $T_e$ where 
the Cooper pairing due to the Kohn-Luttinger (KL) effect will take place.
Below this point there will be an additional NG boson  
contributing to the expression for specific heat. The latter will 
still scale as  $T^3$. Above or below the temperature of the 
KL transition, however, we expect that the 
heat transfer in deuteron condensate will  be dominated  
by temperature-waves ("second sound") in the two-component 
phonon-superfluid medium.

\vspace{0.1in}

The densities of interest in the present work are expected to be present 
in certain astrophysical systems, namely in sub-stellar objects with masses low 
enough not to be able to fuse deuterium. An example is a low-temperature ($T\lsim 10^5~K$),
low-mass ($ M < 13\cdot M_{\text{Jupiter}}$) brown dwarf. 
Such an object is supported by the electron  Fermi degeneracy pressure 
working against gravity. An average interior density in the core 
could be close to $\sim  10^3~g/cm^3$ and the radius -- of the 
order of  Jupiter's radius\footnote{For simplicity, we consider a homogeneous interior. 
This is a good approximation up to a factors of a few, as one can show, 
based on a nonrelativistic polytrope model \cite {Shapiro}.}. 
The abundances of  various nuclei (protons, deuterons and helium) 
in the core of such a dwarf are expected to trace the cosmological abundances of 
the corresponding chemical elements.  In particular,  the fractions  
of protons, deuterons and helium nuclei in the core should be  
similar to the fractions of the corresponding chemical elements   
in Jupiter's core $N_{protons}\simeq 30\cdot N_{helium} \simeq 10^5\cdot N_{deuteron}$. 
Furthermore, we assume a non-convective interior, and the chemical 
stratification  of the elements (separation of chemical elements)
due to rotation of the object.  Under these conditions the very center of 
the core would  dominantly consist of helium sub-core of radius  
$\sim R_{\text{Jupiter}}/2$, surrounded by a thin deuteron 
layer of $\sim 2~km$ width, which is topped by a thick layer 
of protons, $R_{proton}\sim R_{\text{Jupiter}}/2$. 
For temperatures around $10^4~K$ and densities at hand, the mean free path of the 
deuteron nuclei could be estimated to be $\sim 30~m$, suggesting that  
the smearing of the thin deuteron layer into the upper proton  and lower helium 
layers is negligible. Also, the thermo- and pycno-nuclear fusion 
rates are small. Hence, we got conditions appropriate for condensation of the 
deuteron nuclei.  The state of the core of this brown dwarf can 
be described as follows: in the sub-core the helium nuclei are 
crystallized; this is followed by a thin layer 
of a deuteron quantum liquid, which is  topped by a gas of protons that is about to 
become quantum at slightly lower temperatures.  Such an arrangement in the core 
should impact  rotational properties of the brown dwarf. 
In particular, the rotation of the proton layer should not be expected 
in general to be in a phase with the rotation of the solid helium core, 
and therefore, certain irregularities in rotation might emerge.

\vspace{0.3cm}

\subsection*{Acknowledgments}

We'd like to thank Yosi Gelfand, Andy Kent  and Arkady Vainshtein for 
useful discussions. The work of GG is supported by NSF grant PHY-0758032.  
LB and DP are supported respectively by MacCracken and James Arthur 
Graduate Fellowships at NYU.

\vspace{0.3cm}

\renewcommand{\theequation}{A-\Roman{equation}}
\setcounter{equation}{0} 

\subsubsection*{Appendix A. Relativistic Description of the Condensate}

For a relativistic description of the system of deuterons in the 
background of electrons, we use an order parameter $V_\mu$ 
(an early treatment is given in \cite{vz}; for modern discussion, see \cite{pr}). 
The action should possess a local $U(1)_{EM}$, as well as Poincare invariances. 
The dynamics of the system is therefore described by the following effective Lagrangian 
\begin{equation}
\mathcal{L}=-\frac{1}{4g^2}F_{\mu \nu}F^{\mu \nu}-\frac{1}{2} 
G^*_{\mu \nu}G^{\mu \nu}+m^2_H
V^*_{\mu} V^{\mu}-J_{\mu} A^{\mu}.
\label{l2}
\end{equation}
Here $G_{\mu \nu}=D_{\mu}V_{\nu}-D_{\nu}V_{\mu}$ is the field strength 
(covariant derivative being defined as $D_\mu=\partial_\mu-iA_\mu$), 
and $J_\mu$ represents the averaged fermionic current. It should be noted, 
that in this notation $A_\mu$ is rescaled by a factor of $g$, as compared 
to the physical normalization of the photon field. 

There exist three additional dimension-four, gauge-invariant operators 
which could contribute to the effective Lagrangian. These are the 
quartic self-couplings and magnetic dipole interactions of the form
\beq
\mathcal{L}'=-\frac{1}{2} \lambda_1(V^*_{\mu} V^{\mu})^2+\frac{1}{2} 
\lambda_2(V^*_{\mu} V_{\nu})^2+i\kappa V^*_{\mu} V_{\nu}F^{\mu\nu}.
\label{lprime}
\eeq
For the purposes of studying the mass spectrum on the condensate, 
these terms are not of great importance and we will ignore them 
within the present section. They are however fully taken into 
account in the results of section 3.

The equations of motion, following from (\ref{l2})
\beq
\frac{1}{g^2}\partial^{\mu}F_{\mu \nu}+i
(V^{*\mu}G_{\mu \nu}-V^{\mu}G^*_{\mu \nu})=J_\nu, 
\qquad D^\mu G_{\mu \nu}+m_H ^2 V_\nu=0,
\label{eqm}
\eeq
possess a solution 
\beq
A_\mu-\partial_{\mu}\alpha=(m_H,\mathbf{0}), \qquad V_{\mu}=(0,e^{i\alpha}\mathbf{v}),
\eeq
where $\mathbf{v}$ is an arbitrary real vector with the norm $v=\sqrt{J_0/2m_H}$ 
and $\alpha$ is the overall phase of the vector $V_\mu$. 
The norm of $\mathbf{v}$ is fixed by the first  Maxwell equation that 
enforces the neutrality of the system in the bulk of the condensate.
Our choice of vacuum corresponds to $\alpha =0$:
\beq
\langle A_\mu\rangle=(m_H,\mathbf{0}), \qquad 
\langle V_{\mu}\rangle =(0,0,0,v)\equiv v_{\mu}.
\label{vac}
\eeq 
This breaks the local $U(1)$ symmetry, giving mass to the photon. 
It also breaks the spatial $O(3)$ group, down to $O(2)$. We should 
therefore expect two Nambu-Goldstone modes in the spectrum, corresponding 
to the latter symmetry breaking. 

Hamiltonian of the system (excluding the contribution of the 
Fermi energy of electrons), on the above background
\beq
\mathcal{H}=-\mathcal{L}=m_HJ_0,
\eeq
reduces to the rest energy of the vector bosons.\\
\\
Let us consider small perturbations of the background (\ref{vac}) in the 
equations of motion (\ref{eqm}). By setting $V_\mu= v_\mu+\sigma_\mu$ 
and $A_\mu=m_H\delta_\mu^0+a_\mu$, one obtains

\beq
\left[\left(\frac{1}{g^2}\Box+2v^2\right)\eta^{\nu \mu}-\frac{1}{g^2}
\partial^\nu \partial^\mu+ 2 v^\nu v^\mu \right] a_\mu + 
\left[ iv^\alpha \partial_\alpha \eta^{\nu \mu}-i
\partial^\nu v^\mu \right](\sigma_\mu-\sigma^* _\mu)\nonumber \\
+[m_H v^\nu \delta^\mu _0-2m_H \delta^\nu _0 v^\mu](\sigma_\mu+\sigma^* _\mu)=0,~
\eeq

\beq
[i v^\alpha \partial_\alpha \eta^{\nu \mu}-iv^\nu \partial^\mu-2mv^\nu \delta_0 ^\mu+m_H \delta^\nu _0 v^\mu]a_\mu+[(\Box-2im_H \partial_0)\eta^{\nu \mu}-\partial^\nu \partial^\mu \nonumber \\
+im_H(\delta^\nu_0 \partial^\mu+ \partial^\nu \delta^\mu _0)+m_H ^2 \delta^\nu _0 \delta^\mu _0]\sigma_\mu=0,~
\eeq  
where $\eta _{\mu \nu}\equiv diag(1,-1,-1,-1)$ and all the derivatives act to the right. In order to obtain the mass 
gaps of perturbations on the condensate, we take the infinite wavelength 
limit ($\partial_i\rightarrow 0$, where $i=1,2,3$). It is straightforward to 
show that the spectrum contains two massless modes, three masive photon 
degrees of  freedom (mass $\sqrt{2}gv$), and three modes of mass $2m_H$ 
which correspond to relatvistic effect of creation of  
boson-antiboson pair in the condesnate.

\renewcommand{\theequation}{B-\Roman{equation}}
\setcounter{equation}{0} 

\subsubsection*{Appendix B. Non-Relativistic Limit}

Restoring the factors of $c$, but still keeping $\hbar=1$, gives the following expression for the action 
\beq
S=\int dtd^3x\mathcal{L}=\int dtd^3x \left(-\frac{1}{4}F_{\mu \nu}F^{\mu \nu}-\frac{1}{2} G^*_{\mu \nu}G^{\mu \nu}+m^2_H c^2
V^*_{\mu} V^{\mu}-\frac{g}{c}J_{\mu} A^{\mu}\right),
\eeq
where the conventional normalization of the photon field is assumed, so that $D_\mu=\partial_\mu-igA_\mu/c$. 
The physical quantities are given by the following expressions
\beq
\mathbf{E}=-\frac{1}{c}\partial_t\mathbf{A}-\nabla A_0, \quad \mathbf{B}=\nabla \times \mathbf{A}, \quad \rho=\frac{J^0}{c}, \quad j^i=J^i,
\eeq
where $\mathbf{E}$ and $\mathbf{B}$ are the electric and magnetic fields respectively, $\rho$ denotes the number density of electrons and $\mathbf{j}$ is the corresponding current. 
The non-relativistic (magnetic) limit is taken by sending $c$ to infinity, in a way that leaves physical quantities finite. This implies the finiteness of $A_0$, $\mathbf{A}/c$ and $J^0/c$, and we neglect all quantities, suppressed by an extra factor of $1/c$. It should be noted, that the magnetic field happens to be enhanced by a factor of $c$.
We switch to non-relativistic description by the 
following rescaling of the phase
\beq
V^\mu=e^{-im_Hcx^0}W^\mu,
\eeq
where $W^\mu$ describes the non-relativistic vector field. The Lagrangian therefore becomes
\begin{equation*}
\mathcal{L}=-\frac{1}{4}F_{\mu \nu}F^{\mu \nu}-\frac{1}{2} \Sigma^*_{\mu \nu}\Sigma^{\mu \nu}+im_HcW^{*i}\Sigma_{i0}-im_HcW^{i}\Sigma^*_{i0}  +m^2_H c^2 W^*_0 W^0-\frac{g}{c}J_{\mu} A^{\mu},
\end{equation*}
where $\Sigma_{\mu\nu}=D_\mu W_\nu-D_\nu W_\mu$. As a next step, we 
take the  $c\rightarrow\infty$ limit keeping  $cW_0$ and $W_j$ finite.
This is reduces to rescaling $W_0$ by a factor of $c$, and after setting $c=1$ 
again, gives
\beq
\mathcal{L}^{NR}=-\frac{1}{4}F_{\mu \nu}F^{\mu \nu}-\frac{1}{2} \Sigma^*_{ij}\Sigma^{ij}+im_HW^{*i}\Sigma_{i0}-im_HW^{i}\Sigma^*_{i0}  +m^2_HW^*_0 W^0-\nonumber \\ -gJ_{\mu} A^{\mu}.\quad
\eeq 
Using the same reasoning, we could also include the rest of dimension-four operators (\ref{lprime}) in the non-relativistic Lagrangian
\beq
(\mathcal{L}+\mathcal{L}')^{NR}=-\frac{1}{4}F_{\mu \nu}F^{\mu \nu}-\frac{1}{2} \Sigma^*_{ij}\Sigma^{ij}+im_HW^{*i}\Sigma_{i0}-im_HW^{i}\Sigma^*_{i0}  +m^2_HW^*_0 W^0-\nonumber \\ -gJ_{\mu} A^{\mu}-\frac{1}{2}\lambda_1(W^{*i}W_i)^2 +\frac{1}{2}\lambda_2(W^{*i}W_j)^2+ig\kappa W^{*i}W^jF_{ij}. \qquad
\eeq
One can integrate out the nondynamical $W_0$ field from the last expression to obtain 
(\ref{nrl1})\footnote{Parameters $\lambda_1$ and $\lambda_2$ are related to the counterparts of (\ref{nrl1}) by the following redefinitions $\lambda_s=\lambda_2,\quad \lambda_q=\lambda_1-\lambda_2$.}, which was  our starting point for constructing the effective 
theory of section 2.   The latter form of the Lagrangian makes the connection 
with Schr\"odinger's equation explicit.

\renewcommand{\theequation}{C-\Roman{equation}}
\setcounter{equation}{0} 
\subsubsection*{Appendix C. Spectrum of Perturbations}

In this appendix we briefly summarize the derivation of the full linear spectrum of perturbations on the deuteron condensate. The equations of motion for the perturbations \{$\sigma_0$, $a_0$, $\sigma_\alpha$, $\sigma_3$, $a_\alpha$, $a_3$\} of the gauge-invariant variables \{$W_0$, $A_0$, $W_\alpha$, $|W_3|$, $A_\alpha$, $A_3$\}, which follow from the Lagrangian (\ref{nrl'}), read as follows:
\begin{align}
-im_H\partial_\alpha\sigma_\alpha-im_H\partial_3\sigma_3-m_Hva_3+m_H^2\sigma_0=0, \label{pert1}\\ 
\frac{1}{g^2}(-\Delta +m_0^2) a_0+\frac{1}{g^2}\partial_0(\partial_\alpha a_\alpha+\partial_3 a_3)+4m_Hv\sigma_3=0,\\
im_H\partial_\alpha\sigma_0+\left[(-\Delta-2im_H\partial_0)\delta_{\alpha\beta}+\partial_\alpha\partial_\beta\right]\sigma_\beta+\partial_\alpha\partial_3
\sigma_3-i(1-\kappa)v\partial_3 a_\alpha-i\kappa v\partial_\alpha a_3+ \nonumber \\
+\lambda_sv^2(\sigma_\alpha-\sigma_\alpha^*)=0, 
\end{align}
\begin{align}
\frac{i}{2}m_H\partial_3(\sigma_0-\sigma^*_0)+(-\Delta+\partial^2_3+2\lambda_q v^2)\sigma_3+\frac{1}{2}\partial_3\partial_\alpha(\sigma_\alpha+\sigma^*_\alpha)-2m_Hva_0=0,\\
-i(1-\kappa)v\partial_3(\sigma_\alpha-\sigma^*_\alpha)-\frac{1}{g^2}\partial_\alpha\partial_0 a_0+(\frac{1}{g^2}\Box+2v^2)a_\alpha+\frac{1}{g^2}\partial_\alpha\partial_\beta a_\beta+\frac{1}{g^2}\partial_\alpha\partial_3 a_3=0, \\
m_Hv(\sigma_0+\sigma^*_0)-i\kappa v\partial_\alpha(\sigma_\alpha-\sigma^*_\alpha)-\frac{1}{g^2}\partial_3\partial_0 a_0+\frac{1}{g^2}(\Box+\partial ^2_3)a_3+ \frac{1}{g^2}\partial_3\partial_\alpha a_\alpha=0.
\label{pert6}
\end{align}
Here $\alpha=1,2$ and $\sigma_3$ is real. To obtain the spectrum of perturbations, we find the eigenvalues of the matrix, corresponding to the linear system (\ref{pert1})-(\ref{pert6}). After a tedious, but straightforward calculation, we find the following expressions for the spectrum
\beq
\omega_1^2= 2 v^2 g^2 \left(1+\frac{\mathbf{k}^4+2 \mathbf{k}^2 v^2 \lambda_q }{8 g^2 m_H^2 v^2+\left(\mathbf{k}^2+2 v^2 \lambda_q \right)
m_0^2}\right),
\nonumber \\
\omega^2_{2,4}=\frac{1}{8 m_H^2}(\k^4 +8g^2 v^2 m_H^2+2 \k^2 (2m _H^2+v^2 \lambda _s) \pm \nonumber \\
((4m_H^2(2g^2 v^2 +\k^2)-2\k^2 v^2 \lambda _s-\k^4)^2+32 g^2 m_H ^2 v^2 \k ^4 (1-\kappa)^2)^{1/2}),
\nonumber \\
\omega^2_{3,5}=\frac{1}{8 m_H ^2}(\k^4 +8g^2 v^2 m_H ^2+2 \k^2 (2m _H ^2+v^2 \lambda _s) \pm \nonumber \\
((4m_H^2(2g^2 v^2 +\k^2)-2\k^2 v^2 \lambda _s-\k^4)^2+32 g^2 m_H ^2 \k^2(\k \cdot \mathbf{v})^2 (1-\kappa)^2)^{1/2}).
\label{zust!}
\eeq
Here, the dispersion relation for the longitudinal polarization of the massive 
photon is given by $\omega^2_1$;  moreover, $\omega^2_{2,3}$ correspond to the 
transverse polarizations of the photon, and $\omega_{4,5}$ are frequencies for  
two NG modes. The expressions for $\omega^2_{3,5}$ depend on the direction 
of propagation. Setting $\kappa\in [0,2]$ and $\lambda_s\geq0$, guarantees the absence of 
instabilities, as can easily be verified.

It should also  be noted, that the propagation of one of the NG  
bosons in the direction  perpendicular to $\mathbf{v}$ is independent 
of the coupling constant $g$  (putting aside the fact that 
$\lambda_s$ itself would be determined by $g$)
\beq
\omega_5^2=\frac{k^4}{4m^2_H}+\lambda_s\frac{\k^2 v^2}{2m^2_H},~~~for~~~ 
\quad \k\perp\mathbf{v}.
\label{d}
\eeq
This curious fact can be understood by a closer inspection of the system (\ref{pert6}). 
Integrating out the auxiliary variable $\sigma_0$,  as well as the nondynamical 
heavy modes $\phi_\alpha\equiv(\sigma_\alpha+\sigma^*_\alpha)/2$, one arrives at a 
simple equation for the light modes 
\beq
\left(-\Delta-\frac{4m^2_H\partial^2_0}{\Delta}+2\lambda_s v^2\right)
\pi_\alpha=(1-\kappa)v (\partial_3 a_\alpha - \partial_\alpha a_3),
\eeq
where $\pi_\alpha\equiv(\sigma_\alpha-\sigma^*_\alpha)/2i$. 
It can be easily seen for instance, that for $\k_2=\k_3=0$ 
(or $\partial_2=\partial_3=0$  in position space) $\pi_2$ represents the 
NG field, with the dispersion relation  precisely given by (\ref{d}), 
while $\pi_1$ mixes with other bosonic fields of the theory, 
leading to the abovementioned asymmetry between the NG bosons.

\end {document}